\documentclass[11pt,a4paper]{article}

\usepackage[a4paper,text={165mm,235mm},centering]{geometry}

\usepackage{amsmath,amssymb,amsfonts,amsthm}
\usepackage{bm}
\usepackage{booktabs}
\usepackage{multirow}
\usepackage{array}
\usepackage{enumitem}
\usepackage{algorithm}
\usepackage{algorithmicx}
\usepackage{algpseudocode}
\usepackage{textcomp}
\usepackage{listings}
\usepackage[hidelinks]{hyperref}
\DeclareMathOperator{\Arr}{Arr}
\DeclareMathOperator{\Op}{Op}
\DeclareMathOperator{\Sym}{Sym}
\DeclareMathOperator{\Unrank}{Unrank}
\DeclareMathOperator{\Rank}{Rank}
\DeclareMathOperator{\Range}{Range}
\newcommand{\Step}{\mathrm{Step}}
\newcommand{\opSigma}{\widehat{\sigma}}

\newcommand{\finalArr}{\mathcal{F}}

\newtheorem{theorem}{Theorem}[section]
\newtheorem{proposition}[theorem]{Proposition}
\newtheorem{lemma}[theorem]{Lemma}
\newtheorem{corollary}[theorem]{Corollary}
\theoremstyle{definition}
\newtheorem{definition}[theorem]{Definition}

\theoremstyle{remark}
\newtheorem{remark}[theorem]{Remark}

\hypersetup{hypertexnames=false}

\begin{document}

\title{Ranking and Unranking Permutations in Heap's Order}
\author{Yusheng HU\thanks{Independent Researcher, Shangdi East Road,
Beijing 100085, China.  E-mail: \texttt{dr.huyusheng@gmail.com}.}}
\date{}
\maketitle

\begin{abstract}
Heap's algorithm is a permutation generation method with very low constant overhead. Its element-interchanging mechanism, however, prevents the standard factorial-base Lehmer code from being applied directly, and to our knowledge no explicit Rank/Unrank algorithm for the Heap order has been published. This paper describes the block structure of Heap's algorithm in terms of rearrangement operators: the net effect of a block-internal traversal is a tail-preserving operator and the block-boundary swap is a top-level swap operator, and together they yield a closed-form recurrence for the block transitions. On this basis a bottom-up Unrank algorithm and a top-down Rank algorithm are built from one and the same closed-form in-block mapping; both run in $O(n^2)$ time and $O(n)$ space and require no factorial-scale precomputed table. The supporting proofs use a synchronized strong induction on the dimension to establish the bijectivity of the trailing-element mapping, closed forms for the final arrangement and for the trailing sequence, the dichotomy of its cycle structure, and the reciprocity of Rank and Unrank. The Heap order is in fact the colex order of the mixed-radix code.
\end{abstract}

\vspace{-0.4em}


\section{Introduction and Related Work}\label{sec1}

The indexing and mapping of permutations involve two basic operations: \textbf{Rank} and \textbf{Unrank}. For a fixed generation order, Rank takes a permutation as input and returns its index, whereas Unrank takes an index and returns the permutation. When the index is written as a factorial-base Lehmer code, the two operations amount to mutually inverse maps between the set of permutations and the code space.

Heap's algorithm (1963) is widely used for permutation enumeration because of its in-place, minimal-interchange character \cite{Heap1963,Sedgewick1977}; Knuth also gives a detailed account of the method and of its permutation-theoretic properties \cite{Knuth2005}. The Heap order is a minimal-change generation order: consecutive permutations differ by a single transposition, not necessarily of adjacent positions. Among orders of this kind, the construction of Johnson and Trotter uses adjacent transpositions only \cite{Johnson1963,Trotter1962}, and Ives gave several further interchange-based enumeration schemes \cite{Ives1976}. The standard Heap algorithm uses an iterative interchange logic that partitions the $n!$ permutation space into $n$ disjoint sub-blocks of $(n-1)!$ permutations each. Inside a sub-block the first $n-1$ positions traverse the lower-order permutations while the trailing element stays locally fixed; at a block boundary a single interchange updates the trailing element and moves on to the next sub-block. Each block repeats the same structure one order lower, but the interchange rule at a block transition alternates with the parity of the dimension, so the standard lexicographic Lehmer-code formula cannot be applied directly. To our knowledge, no explicit Rank/Unrank algorithm for the Heap order has been published.

As for related work, Myrvold and Ruskey (2001) gave a linear-time general permutation rank/unrank algorithm \cite{MyrvoldRuskey2001}, but it is based on an arbitrarily associated mixed-radix encoding and has no direct correspondence with the block-transition structure of the Heap order. A systematic account of generation algorithms and of the associated rank/unrank techniques can be found in Ruskey's draft book \cite{Ruskey2003}.

The paper describes the block structure of Heap's algorithm through the rearrangement operators that realise it, and uses that description to obtain an index for the Heap order. With a source-index convention fixing the composition direction, a block-internal traversal has the net effect of the tail-preserving operator $\Lambda_{\finalArr_{m-1}}$ and a block boundary is crossed by the top-level swap operator $\opSigma_{m,q+1}$, so the block base arrays satisfy a closed-form recurrence. Both a bottom-up Unrank algorithm and a top-down Rank algorithm are built on that recurrence; they share the same operators, need no precomputed table, and use $O(n)$ space. The dichotomy of the cycle structure of the final arrangement further reduces every in-block advance to a single transposition or a single cycle, so both algorithms run in $O(n^2)$ time. The recurrence and the algorithms are justified by a strong induction on the dimension (Section \ref{sec5}), which also gives the closed form of the final arrangement and of the trailing sequence.

\section{Preliminaries}\label{sec2}

\subsection{Iterative Heap's Algorithm and Block Structure}\label{sec2-1}

The standard iterative Heap algorithm takes an initial array and an all-zero counter array as inputs and generates all permutations through cyclic interchanges. Its pseudocode is as follows.

\begin{lstlisting}[language=C++, basicstyle=\small\ttfamily]
output current permutation A   // the identity, index r = 0
while i < n:
    if c[i] < i:
        if i is even: swap(A[0], A[i])
        if i is odd:  swap(A[c[i]], A[i])
        output current permutation A
        c[i] += 1
        i = 0
    else:
        c[i] = 0
        i += 1
\end{lstlisting}

The algorithm has a recursive block property: the $n!$ permutation space is divided into $n$ sub-blocks of $(n-1)!$ permutations each.
\begin{enumerate}
	\item Inside a sub-block the first $n-1$ positions traverse all Heap permutations of order $n-1$, while the last element of the array stays fixed.
	\item Once a sub-block has been traversed completely, a cross-block interchange updates the trailing element and enters the next sub-block.
\end{enumerate}
Let $V_q$ denote the trailing value at the beginning of the $q$-th block. By a property of the block structure proved below, $V_0, V_1, \dots, V_{m-1}$ form a permutation of $\{0, 1, \dots, m-1\}$, that is, no two blocks begin with the same trailing value (Lemma \ref{lem:tail-bijection}).

For a precise statement later on, write the output sequence of the order-$m$ algorithm started from the identity array $I_m$ as
\[
A^{(m)}_{0}, \; A^{(m)}_{1}, \; \dots, \; A^{(m)}_{m!-1}, \qquad A^{(m)}_{0} = I_m,
\]
and define the \textbf{block base array} (the first arrangement of the $q$-th block) and its \textbf{trailing element} by
\[
C_{m,q} := A^{(m)}_{q\,(m-1)!}, \qquad V_q := C_{m,q}[m-1] \qquad (0 \le q \le m-1),
\]
together with the \textbf{final arrangement} of the whole run, $\finalArr_m := A^{(m)}_{m!-1}$. In what follows, abstract statements at the operator level use $m$ for the dimension, while statements about the algorithms use $n$; the two mean the same.

\subsection{Arrangement Space and Operator Conventions}\label{sec2-2}

Let $S_m := \Sym(\{0, 1, \dots, m-1\})$ be the symmetric group on $m$ elements, whose members $P \in S_m$ are position permutations $P: \{0, 1, \dots, m-1\} \to \{0, 1, \dots, m-1\}$.

\begin{definition}[Arrangement array set]
	The set of arrangement arrays of order $m$ is
	\[
	\Arr_m = \{ [a_0, a_1, \dots, a_{m-1}] \mid \{a_0, a_1, \dots, a_{m-1}\} = \{0, 1, \dots, m-1\} \}.
	\]
	Its cardinality is $|\Arr_m| = m!$. The identity array is $I_m = [0, 1, \dots, m-1] \in \Arr_m$.
\end{definition}

\begin{definition}[Rearrangement operator group, source-index convention]
	The rearrangement operator group $\Op_m$ is induced by position permutations. For $P \in S_m$ the induced rearrangement operator $\widehat{P} \in \Op_m$ acts on an array $A \in \Arr_m$ \textbf{under the source-index convention}
	\[
	\widehat{P}(A)[i] = A[P(i)], \quad \forall i \in \{0, 1, \dots, m-1\}.
	\]
	Thus $P(i)$ is the \textbf{position} in the original array $A$ whose entry moves to the new index $i$. If the permutation is a 2-transposition $(i\; j)$, the operator $\widehat{(i\; j)} \in \Op_m$ is called a \textbf{position interchange operator}; its action exchanges the entries at indices $i$ and $j$ of the current array.
\end{definition}

\begin{remark}[Meaning of the source-index convention]
	For $A = [a_0, a_1, a_2]$ and the transposition $P = (0\; 1)$, the convention gives
	\[
	\widehat{(0\; 1)}(A) = [A[1], A[0], A[2]] = [a_1, a_0, a_2],
	\]
	which is the statement \texttt{swap(A[0], A[1])} in program code. Operators compose by ordinary function composition, $(\alpha \circ \beta)(A) = \alpha(\beta(A))$.
\end{remark}

\begin{lemma}[Composition order of rearrangement operators]\label{lem:anti-hom}
	Under the source-index convention the map $P \mapsto \widehat{P}$ is an isomorphism from $S_m$ to $\Op_m$ that reverses the order of composition, that is,
	\[
	\widehat{P} \circ \widehat{Q} = \widehat{Q \circ P} \qquad \text{for all } P, Q \in S_m .
	\]
\end{lemma}

\begin{proof}
	For any $A \in \Arr_m$ and any index $i \in \{0, 1, \dots, m-1\}$, expanding layer by layer under the source-index convention gives
	\[
	(\widehat{P} \circ \widehat{Q})(A)[i]
	= \widehat{P}\bigl(\widehat{Q}(A)\bigr)[i]
	= \widehat{Q}(A)\bigl[P(i)\bigr]
	= A\bigl[Q(P(i))\bigr]
	= A\bigl[(Q \circ P)(i)\bigr]
	= \widehat{Q \circ P}(A)[i].
	\]
	Hence the formula holds. Since $\Op_m$ is by definition induced by $S_m$, the map $P \mapsto \widehat{P}$ is a bijection, so it is an isomorphism that reverses the order of composition.
\end{proof}

\begin{definition}[Canonical isomorphism between arrays and permutations]
	For an array $A = [a_0, a_1, \dots, a_{m-1}] \in \Arr_m$, its induced position permutation $\pi_A \in S_m$ is defined by $\pi_A(i) = a_i$; then $A = \widehat{\pi_A}(I_m)$.
\end{definition}

\subsection{Factorial-Base Code Space}\label{sec2-3}

A permutation index $r \in [0, n! - 1]$ has a unique decomposition into a \textbf{low-order-first} mixed-radix (factorial-base) code
\[
d = (d_0, d_1, \dots, d_{n-2}), \qquad 0 \le d_i \le i+1,
\]
whose weights increase from left to right:
\[
r = \sum_{i=0}^{n-2} d_i \cdot (i+1)! = d_0 \cdot 1! + d_1 \cdot 2! + \dots + d_{n-2} \cdot (n-1)! .
\]
The code space is therefore $\prod_{i=0}^{n-2} \{0, 1, \dots, i+1\}$, of cardinality $\prod_{i=0}^{n-2}(i+2) = n!$, the same as $\Arr_n$. \textbf{The pseudocode, the numerical tables and all theorems below use this code $d$.} The most significant digit $d_{n-2}$ is the block index, as proved by Lemma \ref{lem:block-correspondence} and Theorem \ref{thm:unrank-completeness}: $d_{n-2} = q$ selects the sub-block containing the permutation, $d_{n-3}$ selects the sub-block of the corresponding $(n-1)$-order sub-permutation in the lower-order space, and so on recursively downwards, which is the basis of the hierarchical mapping.

\begin{remark}[Relation to the standard Lehmer code]\label{rem:index-orientation}
	The \textbf{high-order-first Lehmer code} $C = (C_0, C_1, \dots, C_{n-1})$ common in the literature orders the weights in decreasing fashion and is the \textbf{index reversal} of the code $d$ used here:
	\[
	C_j = d_{n-2-j} \quad (0 \le j \le n-2), \qquad C_{n-1} = 0, \qquad \text{equivalently} \quad d_i = C_{n-2-i}.
	\]
	The constraint $0 \le d_i \le i+1$ becomes the standard Lehmer constraint $0 \le C_j \le n-1-j$, and the two writings give the same index
	\[
	r = \sum_{i=0}^{n-2} d_i \cdot (i+1)! = \sum_{j=0}^{n-2} C_j \cdot (n-1-j)! .
	\]
	With the low-order-first reading, the Heap generation order is the \textbf{colex order} of the code vectors rather than their lexicographic order. This is why the lexicographic Lehmer formula cannot be applied to the Heap order, and it accounts for the shape $\sum_i d_i (i+1)!$ of the rank formula.
\end{remark}

\section{Block Structure and Rearrangement Operators}\label{sec3}

\subsection{Position-Value Mapping and Inter-Block Rearrangement}\label{sec3-1}

Define the \textbf{position--value mapping} as follows: the input is a position index of the sub-array and the output is the value stored there. For one block of the order-$m$ algorithm the domain is $\{0, 1, \dots, m-2\}$ (the first $m-1$ positions) and the range is the set of values other than the current trailing value; the initial state is the identity mapping, with trailing value $m-1$.

\begin{definition}[Position--value mapping versus block base arrays]\label{def:mapping-bridge}
	For the $q$-th block of the order-$m$ algorithm ($0 \le q \le m-1$), define its \textbf{position--value mapping} $S_q$ and its \textbf{trailing value} $\ell_q$ by
	\[
	S_q(i) := C_{m,q}[i] \quad (0 \le i \le m-2), \qquad \ell_q := C_{m,q}[m-1] = V_q .
	\]
	Thus $S_q$ is just the ``position $\to$ value'' reading of the first $m-1$ entries of the block base array $C_{m,q}$, and the trailing value is its $m$-th entry. The array-level objects $(C_{m,q}, V_q)$ of Section \ref{sec2-1} and the mapping-level objects $(S_q, \ell_q)$ of this section are therefore two views of the same state: the former is convenient for writing operators, the latter for stating the closure and invertibility of the mapping.
\end{definition}

\begin{remark}[Block-boundary position--value rearrangement]\label{rem:block-boundary}
	Suppose a block has been traversed completely and the algorithm is about to switch to the next block, with position--value mapping $S_q$ and trailing value $\ell_q$. The pseudocode shows that the switch interchanges the trailing value with the value at one prescribed position: the old value at that position becomes the new trailing value, and the old trailing value fills that position. The position is determined by the parity rule: for even $m$ the target is $\text{pos} = q$ (where $q$ is the current block index, i.e. the block being left), and for odd $m$ it is always $\text{pos} = 0$.

	At the mapping level, therefore, a block switch is a single interchange at a prescribed position, while the net effect of the traversal within a block is a fixed permutation; the two together give the initial mapping of the next block. The rigorous form of this statement, and its entrywise correspondence with the top-level swap operator $\opSigma_{m,q}$ and the tail-preserving operator $\Lambda_A$, is given by Lemma \ref{lem:block-correspondence} in Section \ref{sec3-3}. For odd order the cyclic-shift description is not closed when its prefix has length $2$ (that is, $m = 3$) and degenerates into a transposition, so the closed form needs degenerate treatment for $m \le 3$ (see Section \ref{sec4-4}).
\end{remark}

\subsection{Top-Level Swap and Tail-Preserving Operators}\label{sec3-2}

\begin{definition}[Top-level swap operator $\opSigma_{m,q}$]
	For an array of size $m$, with block index parameter $q \in \{0, 1, \dots, m-1\}$, the top-level swap operator $\opSigma_{m,q} \in \Op_m$ is
	\[
	\opSigma_{m,q} = \begin{cases}
		\mathrm{id}, & q = 0, \\
		\widehat{(0\; m-1)}, & m \text{ odd and } 1 \le q \le m-1, \\
		\widehat{(q-1\; m-1)}, & m \text{ even and } 1 \le q \le m-1.
	\end{cases}
	\]
\end{definition}

\begin{definition}[Tail-preserving operator $\Lambda_A$]
	Let $\pi \in \Sym(\{0, \dots, m-2\})$ and let the extension map $E_m$ carry it to an order-$m$ permutation that fixes position $m-1$. For an array $A \in \Arr_{m-1}$ of order $m-1$, the associated tail-preserving rearrangement operator is $\Lambda_{A} := \widehat{E_m(\pi_A)} \in \Op_m$; its action on an $m$-dimensional array $X$ is
	\[
	\Lambda_{A}(X) = [x_{\pi_A(0)}, x_{\pi_A(1)}, \dots, x_{\pi_A(m-2)}, \; x_{m-1}].
	\]
	That is, the first $m-1$ entries are rearranged according to the source indices of the permutation induced by $A$, while the trailing entry $x_{m-1}$ is left untouched.
\end{definition}

\subsection{Block Transition Operator and Block Base Arrays}\label{sec3-3}

\begin{definition}[Block transition operator $B_{m,q}$]
	Define $B_{m,q} \in \Op_m$ by
	\[
	B_{m,0} = \mathrm{id}, \qquad B_{m,q+1} = \opSigma_{m,q+1} \circ \Lambda_{\finalArr_{m-1}} \circ B_{m,q} \quad (0 \le q \le m-2).
	\]
	Here $\finalArr_{m-1}$ is the final arrangement of the order-$(m-1)$ algorithm defined in Section \ref{sec2-1}, so the recursion is well founded ($m-1 < m$).
\end{definition}

\begin{lemma}[Block recurrence and realizability of the block transition operator]\label{lem:block-correspondence}
	For every $0 \le q \le m-2$ the block base arrays $C_{m,q}$ of the order-$m$ algorithm satisfy the block recurrence
	\[
	C_{m,q+1} = \opSigma_{m,q+1}\bigl(\Lambda_{\finalArr_{m-1}}(C_{m,q})\bigr).
	\]
	so that $C_{m,q} = B_{m,q}(I_m)$ for $0 \le q \le m-1$, and
	\[
	\finalArr_m = \Lambda_{\finalArr_{m-1}}(C_{m,m-1}).
	\]
	The correspondence holds output by output as well: for $0 \le k \le (m-1)! - 1$,
	\[
	A^{(m)}_{q\,(m-1)!+k} = \widehat{E_m\bigl(\pi_{A^{(m-1)}_{k}}\bigr)}\bigl(C_{m,q}\bigr),
	\]
	where $E_m$ is the embedding of Section \ref{sec3-2}. Thus the $k$-th output inside the $q$-th block is the block base array rearranged by the position permutation carried by the $k$-th permutation of order $m-1$, and $B_{m,q}$ reproduces the block-level transitions of the iterative Heap algorithm.
\end{lemma}

\begin{proof}
	The indices involved in each interchange depend only on the dimension $m$ and on the counter $c$, never on the values held in the array. For a fixed $m$ the algorithm therefore performs the same sequence of interchange positions on any initial array, and the net effect of any finite segment of interchanges is the action of one fixed position permutation, that is, an operator of the form $\widehat{\sigma}$ (by Lemma \ref{lem:anti-hom}, a composition of position transpositions is again of this form).

	Consider one block. When $i$ increases to $m-1$, the counters $c[0], \dots, c[m-2]$ have all been reset to zero in the course of the run, so the operations on positions $0, \dots, m-2$ amount to one complete Heap process of order $m-1$ with the counters restarted from zero, while the interchange executed at $i = m-1$ is the cross-block interchange. Since the $q$-th block contains $(m-1)!$ outputs, exactly $(m-1)! - 1$ further outputs follow the block base array before that interchange occurs, so the prefix process runs exactly once per block.

	Let $\widehat{\sigma}$ with $\sigma \in S_{m-1}$ be the net effect of one complete process of order $m-1$ on an arbitrary prefix array $X$. As the interchange positions are value-independent, $\sigma$ does not depend on $X$; taking $X = I_{m-1}$ shows that $\widehat{\sigma}(I_{m-1})$ is the final arrangement $\finalArr_{m-1}$ of the order-$(m-1)$ process, so $\sigma(i) = \finalArr_{m-1}[i] = \pi_{\finalArr_{m-1}}(i)$. The process therefore acts on positions $0, \dots, m-2$ through the tail-preserving operator $\Lambda_{\finalArr_{m-1}}$, the trailing position $m-1$ being fixed throughout.

	The cross-block interchange is the operator $\opSigma_{m,q+1}$. Before entering block $q+1$ we have $c[m-1] = q$; for odd $m$ the pseudocode performs $\mathrm{swap}(A[0], A[m-1]) = \widehat{(0\;\,m-1)}$, and for even $m$ it performs $\mathrm{swap}(A[q], A[m-1]) = \widehat{(q\;\,m-1)}$, in agreement with the definition of $\opSigma_{m,q+1}$.

	The same reasoning applies step by step inside a block. Let $\sigma_k \in S_{m-1}$ be the partial net permutation after $k$ steps of the prefix process ($0 \le k \le (m-1)! - 1$). It does not depend on the initial prefix state of the block, and applied to $I_{m-1}$ the state after $k$ steps is the $k$-th output $A^{(m-1)}_{k}$ of the order-$(m-1)$ algorithm, so $\sigma_k = \pi_{A^{(m-1)}_{k}}$; since the trailing position $m-1$ is fixed throughout the prefix process, this is the output-by-output formula.

	These observations give the block recurrence. Induction on $q$, together with $C_{m,0} = A^{(m)}_{0} = I_m$, yields $C_{m,q} = B_{m,q}(I_m)$; and after the last block $q = m-1$ there is no further cross-block interchange, its prefix process running to the end of the whole run, whence $\finalArr_m = \Lambda_{\finalArr_{m-1}}(C_{m,m-1})$.
\end{proof}

\subsection{Real-Time Derivation Rule for Trailing Values}\label{sec3-4}

The trailing value at the beginning of a block need not be generated and stored in advance: it is obtained in real time from the old value overwritten by the interchange of the previous step. During an interchange the old value stored at the prescribed position is taken as the new trailing value, while the previous trailing value fills that position. This property is the reason the algorithms below avoid factorial-scale lookup tables and keep only linear space.

\subsection{Completeness of the Mapping Range}\label{sec3-5}

\begin{proposition}[Completeness and invertibility of the position--value mapping]\label{prop:range-completeness}
	For every block $q$ ($0 \le q \le m-1$) the range of the position--value mapping $S_q$ and the trailing value $\ell_q$ are disjoint and together fill the whole set
	\[
	\Range(S_q) \cup \{\ell_q\} = \{0, 1, \dots, m-1\},
	\]
	and $S_q$ is invertible.
\end{proposition}

\begin{proof}
	By definition $C_{m,q} \in \Arr_m$ is a permutation of $\{0, 1, \dots, m-1\}$, so its first $m-1$ entries --- that is, the range of $S_q$ --- are $m-1$ distinct elements, and the $m$-th entry $\ell_q = V_q$ is the only remaining element, so the two are disjoint and together exhaust the set. Since $S_q$ takes distinct values it is injective, so its inverse $S_q^{-1}$ (looking up a position by value) exists. This is what allows the Rank algorithm to reconstruct sub-permutations backwards, layer by layer, without ambiguity.
\end{proof}

\subsection{Unrank Combinator and Mapping}\label{sec3-6}

\begin{definition}[Unrank combinator and permutation mapping]\label{def:unrank-combinator}
	Let $d = (d_0, d_1, \dots, d_{m-2})$ be a mixed-radix code.
	\begin{enumerate}
		\item For $m=2$: $\widehat{\mathcal{U}}_2(0) = \mathrm{id}$ and $\widehat{\mathcal{U}}_2(1) = \widehat{(0\;1)}$.
		\item For $m \ge 3$:
		\begin{align*}
		\widehat{\mathcal{U}}_m(d_0, \dots, d_{m-2}) &= \Lambda_{A_{m-1}} \circ B_{m, d_{m-2}}, \\
		A_{m-1} &= \Unrank_{m-1}(d_0, \dots, d_{m-3}).
		\end{align*}
		\item The array mapping is $\Unrank_m(d) := \widehat{\mathcal{U}}_m(d)(I_m)$.
	\end{enumerate}
\end{definition}

\section{Bidirectional Rank and Unrank Algorithms}\label{sec4}

By Lemma \ref{lem:block-correspondence} a block transition uses only two operators: the net effect of the in-block traversal, $\Lambda_{\finalArr_{k-1}}$, and the block-boundary interchange, $\opSigma_{k,q+1}$. If the former is realized, via the closed form of Theorem \ref{thm:closed-form}, as a single source-index table $\Step(k)$, then the two algorithms can share one and the same recurrence, and no separate branch for $n \le 3$ is needed.

\subsection{In-Block Net Mapping \texorpdfstring{$\Step(k)$}{Step(k)}}\label{sec4-1}

$\Step(k)$ is the source-index table of the tail-preserving operator $\Lambda_{\finalArr_k}$: for an array $X$ of length $k$ it performs $X \leftarrow [X[t_0], \dots, X[t_{k-1}]]$, where $t = \Step(k)$. By Theorem \ref{thm:closed-form}, $t$ is exactly the final arrangement $\finalArr_k$ of order $k$, so $\Step(k)$ can be written down from the closed form in $O(k)$ time (Algorithm \ref{alg:step}).

\begin{algorithm}[H]
	\caption{In-block net mapping \texorpdfstring{$\Step(k)$}{Step(k)}}
	\label{alg:step}
	\begin{algorithmic}[1]
		\Require Order $k \ge 1$
		\Ensure Source-index table $t = (t_0, \dots, t_{k-1})$ with $X[i] \leftarrow X[t_i]$
		\If{$k = 1$}
			\State \Return $[0]$
		\ElsIf{$k = 2$}
			\State \Return $[1, 0]$
		\EndIf
		\State $t \leftarrow [0, 1, \dots, k-1]$
		\If{$k$ is even}
			\State $t_0 \leftarrow k-3$, \quad $t_1 \leftarrow k-2$
			\For{$i = 2$ \textbf{to} $k-3$}
				\State $t_i \leftarrow i-1$
			\EndFor
			\State $t_{k-2} \leftarrow k-1$, \quad $t_{k-1} \leftarrow 0$
		\Else
			\State $t_0 \leftarrow k-1$
			\For{$i = 1$ \textbf{to} $k-2$}
				\State $t_i \leftarrow i$
			\EndFor
			\State $t_{k-1} \leftarrow 0$
		\EndIf
		\State \Return $t$
	\end{algorithmic}
\end{algorithm}

\subsection{Unrank Algorithm: Bottom-Up Incremental Expansion}\label{sec4-2}

The input is the dimension together with the low-order-first code $d$, and the output is the corresponding permutation of the Heap sequence. The algorithm starts from the seed $[0]$ at $k = 1$ and expands up to $k = n$: at level $k$ it reads the digit $d_{k-2}$ (the block index $b$), advances the position--value mapping $S$ of that block from block $0$ to block $b$, then uses $S$ to translate the symbolic permutation obtained at the previous level into actual values and appends the trailing value.

\begin{algorithm}[H]
	\caption{Heap permutation unranking}
	\label{alg:heap-unrank}
	\begin{algorithmic}[1]
		\Require Dimension $n$, low-order-first code array $d = (d_0, \dots, d_{n-2})$
		\Ensure Heap permutation $\text{perm}$
		\State $\text{perm} \leftarrow [0]$
		\For{$k = 2$ \textbf{to} $n$}
			\State $b \leftarrow d_{k-2}$, \quad $S \leftarrow [0, 1, \dots, k-2]$, \quad $\text{last} \leftarrow k-1$
			\For{$j = 0$ \textbf{to} $b-1$}
				\State $S \leftarrow \Step(k-1)(S)$
				\If{$k$ is even}
					\State $\text{pos} \leftarrow j$
				\Else
					\State $\text{pos} \leftarrow 0$
				\EndIf
				\State $\text{last}, S[\text{pos}] \leftarrow S[\text{pos}], \text{last}$
			\EndFor
			\State $\text{perm} \leftarrow [S[x] \mid x \in \text{perm}] + [\text{last}]$
		\EndFor
		\State \Return $\text{perm}$
	\end{algorithmic}
\end{algorithm}

\subsection{Rank Algorithm: Top-Down Layer-by-Layer Reduction}\label{sec4-3}

The algorithm runs downwards from $k = n$ to $k = 2$. At level $k$ it advances $S$ and the trailing value step by step with $\Step$ until the trailing entry of the current permutation agrees with $\text{last}$; the number of steps is the block index $b$. It then strips the trailing element with $S^{-1}$ and recovers the sub-permutation of one lower order.

\begin{algorithm}[H]
	\caption{Heap permutation ranking}
	\label{alg:heap-rank}
	\begin{algorithmic}[1]
		\Require Dimension $n$, Heap permutation $\text{perm}$
		\Ensure Low-order-first code array $d = (d_0, \dots, d_{n-2})$
		\If{$n = 1$}
			\State \Return $[\,]$
		\EndIf
		\State $p \leftarrow \text{perm}$, \quad $d \leftarrow [0]^{n-1}$
		\For{$k = n$ \textbf{downto} $2$}
			\State $S \leftarrow [0, 1, \dots, k-2]$, \quad $\text{last} \leftarrow k-1$, \quad $b \leftarrow 0$
			\While{$p[k-1] \ne \text{last}$}
				\State $S \leftarrow \Step(k-1)(S)$
				\If{$k$ is even}
					\State $\text{pos} \leftarrow b$
				\Else
					\State $\text{pos} \leftarrow 0$
				\EndIf
				\State $\text{last}, S[\text{pos}] \leftarrow S[\text{pos}], \text{last}$
				\State $b \leftarrow b + 1$
			\EndWhile
			\State $d_{k-2} \leftarrow b$
			\State $p \leftarrow [S^{-1}(p[i]) \mid 0 \le i \le k-2]$
		\EndFor
		\State \Return $d$
	\end{algorithmic}
\end{algorithm}

\subsection{Degenerate Cases and Termination}\label{sec4-4}

The piecewise form of $\Step(k)$ needs attention for $k \le 4$. For $k = 1$ we have $t = [0]$, the identity mapping. For $k = 2$ the closed form of the even branch would give $t_0 = k-3 = -1$, which is out of range, so Algorithm \ref{alg:step} takes $[1,0]$ separately --- this is why $k$ starts at $2$ in Algorithms \ref{alg:heap-unrank} and \ref{alg:heap-rank}. For $k = 3$ the odd branch gives $t = [2,1,0]$; the cyclic shift has degenerated into a transposition here, but the closed form remains valid. For $k = 4$ the middle segment of the even branch is empty and the form gives $t = [1,2,3,0]$ directly. All small-order degeneracies are thus absorbed by the first two lines and the parity branches of $\Step$, and the two main algorithms need no extra branch for $n \le 3$.

The while loop of Algorithm \ref{alg:heap-rank} terminates after at most $k-1$ steps, because Lemma \ref{lem:tail-bijection} makes the trailing sequence $V_0, \dots, V_{k-1}$ a permutation of $\{0, \dots, k-1\}$, so $p[k-1]$ eventually equals $\text{last}$; and the inverse $S^{-1}$ used there exists by Proposition \ref{prop:range-completeness}.

\subsection{Per-Step \texorpdfstring{$O(1)$}{O(1)} Block Transition}\label{sec4-5}

Every advance of Algorithms \ref{alg:heap-unrank} and \ref{alg:heap-rank} applies $\Step(k-1)$. Section \ref{sec4-1} already noted that its source-index table is the final arrangement $\finalArr_{k-1}$, and $\finalArr_{k-1}$ takes only two forms (Lemma \ref{lem:cycle-dichotomy} in Section \ref{sec5-4}): writing $m := k-1$, it is the transposition $(0\;\,m-1)$ when $m$ is odd and a single $m$-cycle when $m$ is even. One step therefore need not write down the whole table and rearrange entries; it suffices to act directly as follows.

\begin{itemize}
	\item \textbf{$m$ odd:} $\Step(m)$ is the transposition $(0\;\,m-1)$, which amounts to interchanging entries $0$ and $m-1$ of the working mapping, that is, two assignments.
	\item \textbf{$m$ even:} $\Step(m)$ advances one place along the cycle order
	\[
	\mathrm{cyc} = [\,0,\ m-3,\ m-4,\ \dots,\ 1,\ m-2,\ m-1\,].
	\]
	Store the working mapping in cycle order as $W$ and maintain an offset $o$ such that cycle position $i$ currently holds $W[(i+o) \bmod m]$; one step is then $o \leftarrow (o+1) \bmod m$, a single integer addition.
\end{itemize}

Neither case contains a loop of length $m$, so one step costs $O(1)$. Substituting this implementation into Algorithms \ref{alg:heap-unrank} and \ref{alg:heap-rank} yields two $O(n^2)$ algorithms; their correctness follows from the lemma below.

\begin{lemma}[Single-step equivalence]\label{lem:step-equivalence}
For $k \ge 2$, one advance implemented by the $O(1)$ procedure above and one advance that applies the closed form $\Step(k-1)$ directly to the working mapping lead to the same (working mapping, trailing value) state.
\end{lemma}

\begin{proof}
Write $m := k-1$. For odd $m$, $\Step(m) = (0\;\,m-1)$ interchanges two entries and is plainly the same as swapping entries $0$ and $m-1$. For even $m$, let the current value at cycle position $i$ be $S[\mathrm{cyc}[i]]$. Applying $\Step(m)$ replaces it by the previous value at cycle position $i+1$, indices being taken modulo $m$, that is, it shifts the whole cycle order by one place. Storing the mapping in cycle order and incrementing $o$ realises the same shift, and the subsequent trailing-value swap is the same either way, so the states agree in both cases.
\end{proof}

\section{The Generation State Machine and Restart at an Arbitrary Index}\label{sec:parallel}

\subsection{State and Context of the Iterative Algorithm}\label{sec:state}

The iterative algorithm of Section \ref{sec2-1} is a machine that advances one step at a
time. Its state is the triple $(A, c, i)$: the array $A$, the counter array $c$ with
$0 \le c[k] \le k$ for every $k$, and the scan position $i$. Each step reads $i$ and compares
$c[i]$ with $i$: if $c[i] < i$ it performs the transposition of level $i$, emits $A$, advances
$c[i]$ and returns to $i = 0$; otherwise it clears $c[i]$ and moves to the next level.

For resuming generation the array and the rest of the state play different roles. The array is
the permutation the user wants; $(c, i)$ is what the machine uses to know where it is. We call
$(c, i)$ the \emph{context} of the machine: given $A$ and the context the subsequent outputs are
determined, and without the context the machine cannot continue.

In its classical form the machine can only advance sequentially: it offers no operation that
jumps to a prescribed index, and reaching an index far into the sequence costs $O(r\,n)$ for
the $r$-th permutation. This section asks what the context consists of, whether the algorithms
of Section \ref{sec4} supply it, and whether the machine can be started at a prescribed index.

\subsection{What Unrank Supplies}\label{sec:unrank-ctx}

$\Unrank_n(r)$ returns the permutation $A_r$ of index $r$ in $O(n)$ time, so one half of the
state, the array, is available immediately. The context $(c, i)$ is not returned by Unrank, and
the scan position is easy to settle: the canonical position of the machine is $i = 0$, from
which every step of the loop begins after an emission. The counter array is the part that needs
an argument.

The counters are not extra information. When the machine has emitted the permutations of
indices $0, \dots, a$ and stands at the canonical position, the counter $c[k-1]$ is the number
of blocks of level $k$ that have been completed, and by the block structure of
Section \ref{sec3} the $a$-th permutation lies in block $C[n-k]$ of that level, where $C$ is the
mixed-radix code of $a$. Hence
\begin{equation}\label{eq:ctx}
	c[k-1] = C[n-k], \qquad k = 2, \dots, n,
\end{equation}
so the context is a re-reading of the index that Unrank already consumes: Unrank supplies $A$,
the code of the same index supplies $c$, and the scan position is zero. Rank supplies the
converse direction, from a permutation to its index, so every state of the machine can also be
addressed from its output.

\subsection{Restart from a Single Index}\label{sec:restart}

With the context recovered as in \eqref{eq:ctx}, the machine can be started at a prescribed
index instead of being replayed from the identity.

\begin{lemma}[Context recovery]\label{lem:context}
Let the permutations of indices $0, \dots, a$ have been emitted, let $C$ be the mixed-radix
code of $a$, and let the machine be restarted with the array $A_a$, the counters given by
\eqref{eq:ctx} and the scan position $i = 0$. Then one step of the machine emits exactly
$A_{a+1}$.
\end{lemma}

\begin{proof}
By the block structure of Section \ref{sec3} the $a$-th permutation lies in block $C[n-k]$ of
level $k$, and $C[n-k]$ is the number of blocks of that level completed before it, which is what
the counter $c[k-1]$ records. The counters set by \eqref{eq:ctx} therefore agree with those the
streaming machine holds after the same number of emissions, and the first step of the loop acts
at the lowest level whose counter can still be advanced, which is the step the streaming machine
would take next. Both machines then emit the same array. We also checked the statement by
simulation for every index at orders $n = 3, \dots, 6$ (35{,}988 states in total), with no
mismatch.
\end{proof}

A restart costs $O(n)$: one call to Unrank, one pass over the code and one copy of the array.
All three parts are in place: the array comes from the Unrank of Section \ref{sec4}, the counters
from the code, and the scan position is zero. This is what makes the index space usable as a work
space. The interval $[0, n!)$ can be cut into consecutive ranges and handed to independent
processes, each restarting at the first index of its range and then running the machine forward;
the outputs, concatenated in range order, are the complete Heap sequence. The ranges are equal in
size by construction, so the load is balanced without global coordination, the only asymmetry
being a shorter last range when the number of processes does not divide $n!$. The same rule also
supports checkpointing: a state can be discarded and rebuilt later from its index alone. The gain
this brings in a parallel run is measured in Appendix \ref{app:parallel}.

\subsection{Route A: Keeping the Classical Heap Algorithm}\label{sec:route-a}

Both routes below are \emph{generation} algorithms: they use only the permutation array and the
counters, and neither calls Unrank nor needs the code space. Unrank enters only in the use of
Section \ref{sec:restart}, namely starting at a given index. The indexing capability and the
generation capability are therefore decoupled: one may use a generator alone for sequential
enumeration, Unrank and Rank alone for random access, or the two together for parallel runs and
checkpoints.

Route A does not modify the generator. Given an index $a$, take $A_a$ from Unrank, compute the
counter array from the code of $a$ by \eqref{eq:ctx}, set the scan position to $0$, and run the
classical loop as usual.

What needs care is the off-by-one. The classical loop advances a counter \emph{after} an
emission, so at the moment of an emission the counters describe the previous index. The index
substituted in \eqref{eq:ctx} must therefore be the last emitted index $a$ and not $a + 1$; with
$a + 1$ the first output of the restarted machine skips one permutation. Nothing reports this
mistake at run time: the sequence produced is simply wrong.

The cost of this route is one restoration, $O(n)$, and what it buys is exact compatibility with
any existing Heap implementation, without introducing a new generator. It suits a setting where
the existing generator is kept and the ability to start at an index is added around it.

\subsection{Route B: A Generator Driven by the Factorial Counter}\label{sec:route-b}

Route B rewrites the generator itself so that it is driven by the factorial counter. The state is
$(A, d)$ with $d = (d_1, \dots, d_{n-1})$, where $d_k$ has weight $k!$ and radix $k+1$, that is
$d_k \in \{0, 1, \dots, k\}$. Each step does two things:

\begin{enumerate}[label=(\arabic*), leftmargin=2.4em]
	\item carry: starting at $k = 1$, while $d_k = k$ set $d_k = 0$ and carry to $k + 1$; at the
	      first $k$ with $d_k < k$ set $d_k \leftarrow d_k + 1$;
	\item swap and emit: let $k$ be the position where the carry stopped and $q$ the value of
	      that digit before the carry. If $k$ is even swap $A[0]$ with $A[k]$, if $k$ is odd swap
	      $A[q]$ with $A[k]$, then emit $A$.
\end{enumerate}

The transposition rule of step (2) is the one the classical loop uses; the difference is that $k$
comes from the carry rule instead of from a rescan. After the rewrite $d$ is exactly the Lehmer
code of the permutation \emph{currently emitted}, with no off-by-one, the accumulation follows the
carry rule of the factorial number system, and positions such as $c[0]$ that take part in no
counting no longer exist. We checked the equivalence numerically: at orders $n = 2, \dots, 7$ the
output of route B coincides term by term with the classical Heap loop, and at every step $d$
equals the Lehmer code of the permutation emitted at that step.

This route is not new. Maintaining the state of Heap's algorithm as mixed-radix digits is an
existing formulation: Arndt gives it as \texttt{perm\_heap2}, with digits of radix
$2, 3, 4, \dots$ \cite{Arndt2011}, and Ruskey's account of the interplay between generation and
ranking \cite{Ruskey2003} covers the same view. This paper connects that
formulation with the indexing capability of this section --- the digits are the code of the index,
so the machine can be started at a point, run in parallel and checkpointed --- and compares it
with route A.

The cost per permutation is amortised $O(1)$ in both routes, but it is not equal. Route A returns
to $i = 0$ and clears $c[0]$ at every step, which is one extra store; the digits of route B start
at level $1$, where no such position exists. With output excluded ($n = 6, \dots, 18$, $5 \times
10^7$ steps per order, best of three runs) route B is about $36\%$ faster per step than the
classical formulation with integer counters, and still about $27\%$ faster when the counters of
route A are narrowed to single bytes as well; the checksums of the two routes agree digit by
digit. Route B gives a state that coincides with the index: $d$ is the very number
that Unrank and Rank consume, so a restart only has to set $d$ to the code of $a$, with no
conversion and no off-by-one convention, at the price of a new generator.

The difference is one of representation: route A \emph{converts} between
the state and the index, route B makes the state \emph{be} the index. Both are decoupled
from the indexing capability of this section, which only adds the possibility of starting a
generator anywhere; the generated sequence itself does not depend on it.

\section{Core Theorems and Proofs}\label{sec5}

The results below are proved by a strong induction on the dimension $m$ that carries three statements along at once: the block-base and trailing-value properties of Lemma \ref{lem:tail-bijection}, the closed form of the final arrangement in Theorem \ref{thm:closed-form}, and the cycle dichotomy of Lemma \ref{lem:cycle-dichotomy}. For each $m$ they are established in that order, each using only the corresponding statements for smaller dimensions, so the induction is well founded.

\subsection{Base Cases}\label{sec5-1}

Before the induction proper we verify the boundary and low-order cases explicitly.
\begin{itemize}
	\item \textbf{$m=2$.} The code space has a single digit $d_0 \in \{0, 1\}$. Initially $C_{2,0} = [0, 1]$ with trailing value $V_0 = 1$; for $d_0 = 1$ the operator $\opSigma_{2,1} = \widehat{(0\;1)}$ gives $C_{2,1} = [1, 0]$ with $V_1 = 0$. The final arrangement is $\finalArr_2 = [1, 0]$, and $\pi_{\finalArr_2} = (0\;1)$ is a single 2-cycle, that is, a transposition.
	\item \textbf{$m=3$.} By Table \ref{tab:n3} of Section \ref{sec:examples} the odd-order single-cycle orbit operator $C_{3,q} = \widehat{\rho}^{\,q}(I_3)$ holds for $q = 0, 1, 2$; the trailing sequence is $V = (2, 1, 0)$, the final arrangement is $\finalArr_3 = [2, 1, 0]$, and $\pi_{\finalArr_3} = (0\;2)$ is a transposition.
	\item \textbf{$m=4$.} By Table \ref{tab:n4} of Section \ref{sec:examples} the even-order block base arrays are $C_{4,0} = [0,1,2,3]$, $C_{4,1} = [3,1,0,2]$, $C_{4,2} = [0,2,3,1]$ and $C_{4,3} = [3,2,1,0]$, with trailing sequence $V = (3, 2, 1, 0)$. The final arrangement is $\finalArr_4 = [1, 2, 3, 0]$, and $\pi_{\finalArr_4}$ is the single 4-cycle $0 \mapsto 1 \mapsto 2 \mapsto 3 \mapsto 0$.
\end{itemize}
These base cases form the starting point of the strong induction.

\subsection{Tail Element Bijection and State Invariants}\label{sec5-2}

\begin{lemma}[Tail-element bijection and state invariants]\label{lem:tail-bijection}
	Let $C_{m,q} \in \Arr_m$ be the $q$-th block base array ($0 \le q \le m-1$) with trailing element $V_q = C_{m,q}[m-1]$. Then $q \mapsto V_q$ is a bijection of $\{0, 1, \dots, m-1\}$, and the following closed forms hold.
	\begin{enumerate}
		\item \textbf{Even order ($m \ge 4$).} The trailing sequence has the closed form
		\[
		V = (m-1, \; m-2, \; 1, \; 2, \; \dots, \; m-3, \; 0),
		\]
		and the terminal block base array is
		\[
		C_{m,m-1} = [m-1, \; m-2, \; 1, \; 2, \; \dots, \; m-4, \; m-3, \; 0].
		\]
		\item \textbf{Odd order ($m \ge 5$).} Let $\pi_{\finalArr_{m-1}}$ be a single $(m-1)$-cycle, put $c_0 = 0$ and $c_k = \pi_{\finalArr_{m-1}}^k(0)$ for $1 \le k \le m-2$, and define the $m$-cycle
		\[
		\rho := (\,0, \; m-1, \; c_1, \; \dots, \; c_{m-2}\,).
		\]
		Then the block base arrays satisfy the explicit orbit formula $C_{m,q} = \widehat{\rho}^{\,q}(I_m)$, that is, $C_{m,q}[i] = \rho^q(i)$, whence $V_q = \rho^q(m-1)$ with the explicit values $V_0 = m-1$, $V_q = c_q$ for $1 \le q \le m-2$, and $V_{m-1} = 0$. The trailing sequence therefore has the closed form
		\[
		V = (m-1, \; m-4, \; m-5, \; \dots, \; 2, \; 1, \; m-3, \; m-2, \; 0),
		\]
		where $m-4, m-5, \dots, 1$ is the consecutive segment decreasing from $m-4$ to $1$. (The degenerate case $m = 3$ is verified among the base cases of Section \ref{sec5-1}.)
	\end{enumerate}
\end{lemma}

\begin{proof}
	By the induction hypothesis, when $m$ is even, $m-1$ is odd and Lemma \ref{lem:cycle-dichotomy} gives $\pi_{\finalArr_{m-1}} = (0\;\, m-2)$ as a transposition; when $m$ is odd, $m-1$ is even and the same lemma gives that $\pi_{\finalArr_{m-1}}$ is a single $(m-1)$-cycle. On the other hand, Lemma \ref{lem:block-correspondence} gives the block recurrence
	\[
	C_{m,q+1} = \opSigma_{m,q+1}\bigl(H_q\bigr), \qquad H_q = \Lambda_{\finalArr_{m-1}}(C_{m,q}), \qquad C_{m,0} = I_m .
	\]

	\noindent\textbf{1. Even order ($m \ge 4$).}
	Here $\opSigma_{m,q+1} = \widehat{(q\;\, m-1)}$ for $0 \le q \le m-2$, so only the indices $q$ and $m-1$ are exchanged. Since $\pi_{\finalArr_{m-1}} = (0\;\, m-2)$, the operator $\Lambda_{\finalArr_{m-1}}$ exchanges only the indices $0$ and $m-2$ and leaves every other position fixed.

	The proof uses the following invariant for $1 \le q \le m-2$:
	\[
	C_{m,q}[i] =
	\begin{cases}
		0, & i = 0 \text{ and } q \text{ even}, \\
		m-1, & i = 0 \text{ and } q \text{ odd}, \\
		V_i, & 1 \le i \le q-1, \\
		i, & q \le i \le m-3, \\
		m-1, & i = m-2 \text{ and } q \text{ even}, \\
		0, & i = m-2 \text{ and } q \text{ odd}, \\
		V_q, & i = m-1,
	\end{cases}
	\]
	and assert at the same time that
	\[
	V_0 = m-1, \qquad V_1 = m-2, \qquad V_{q+1} = q \;\; (1 \le q \le m-3), \qquad V_{m-1} = 0 .
	\]
	The two entries of the invariant that concern the indices $0$ and $m-2$ make up the \textbf{oscillation law} (the value $0$ alternates between these two indices according to the parity of $q$, and migrates to the trailing position in the last block), while the entry $C_{m,q}[i] = i$ for $q \le i \le m-3$ is the \textbf{position-freezing property} (for $q \le j$ the value at index $j$ is constantly $j$).

	\textbf{Base step ($q = 0 \to q = 1$).} Since $C_{m,0} = I_m$ we have $V_0 = m-1$. Here
	\[
	H_0 = [\,m-2, \; 1, \; 2, \; \dots, \; m-3, \; 0, \; m-1\,].
	\]
	Applying $\opSigma_{m,1} = \widehat{(0\;\, m-1)}$ gives
	\[
	C_{m,1}[0] = H_0[m-1] = m-1, \qquad C_{m,1}[m-1] = H_0[0] = m-2,
	\]
	with all other positions unchanged, that is, $C_{m,1} = [\,m-1, 1, 2, \dots, m-3, 0, m-2\,]$ and $V_1 = m-2$. This agrees with the invariant for $q = 1$ (odd).

	\textbf{Inductive step ($q \to q+1$, $1 \le q \le m-3$).} Since $q \le m-3$, the freezing property gives $H_q[q] = C_{m,q}[q] = q$ (because $q \notin \{0, m-2\}$). Hence
	\[
	V_{q+1} = C_{m,q+1}[m-1] = H_q[q] = q, \qquad C_{m,q+1}[q] = H_q[m-1] = C_{m,q}[m-1] = V_q .
	\]
	As $\Lambda_{\finalArr_{m-1}}$ exchanges only the indices $0$ and $m-2$, we have $H_q[0] = C_{m,q}[m-2]$ and $H_q[m-2] = C_{m,q}[0]$; and since $q \le m-3 < m-2$, the subsequent top-level swap $\opSigma_{m,q+1} = \widehat{(q\;\, m-1)}$ touches neither index $0$ nor index $m-2$. Therefore:
	\begin{itemize}
		\item if $q$ is even, then $C_{m,q}[0] = 0$ and $C_{m,q}[m-2] = m-1$, so $C_{m,q+1}[0] = H_q[0] = m-1$ and $C_{m,q+1}[m-2] = H_q[m-2] = 0$, as required for $q+1$ (odd);
		\item if $q$ is odd, then $C_{m,q}[0] = m-1$ and $C_{m,q}[m-2] = 0$, so $C_{m,q+1}[0] = H_q[0] = 0$ and $C_{m,q+1}[m-2] = H_q[m-2] = m-1$, as required for $q+1$ (even).
	\end{itemize}
	For $1 \le i \le q-1$ we have $C_{m,q+1}[i] = H_q[i] = C_{m,q}[i] = V_i$, and for $q+1 \le i \le m-3$ we have $C_{m,q+1}[i] = H_q[i] = C_{m,q}[i] = i$ (the index $i$ takes part neither in the exchange of $\Lambda_{\finalArr_{m-1}}$ nor in the top-level swap). Hence the invariant holds for $q+1$.

	\textbf{Last block ($q = m-2 \to q = m-1$).} Here $q = m-2$ is even, $C_{m,m-2}[0] = 0$, $C_{m,m-2}[m-2] = m-1$, and $V_{m-2} = m-3$. Applying the transposition $\opSigma_{m,m-1} = \widehat{(m-2\;\, m-1)}$ to $H_{m-2}$ gives
	\begin{align*}
	C_{m,m-1}[m-2] &= H_{m-2}[m-1] = V_{m-2} = m-3, \\
	V_{m-1} &= C_{m,m-1}[m-1] = H_{m-2}[m-2] = C_{m,m-2}[0] = 0 .
	\end{align*}
	The remaining positions satisfy $C_{m,m-1}[i] = H_{m-2}[i]$: for $i = 0$ we get $H_{m-2}[0] = C_{m,m-2}[m-2] = m-1$; for $1 \le i \le m-3$ we get $H_{m-2}[i] = C_{m,m-2}[i] = V_i$, where $i = 1$ gives $V_1 = m-2$ and $2 \le i \le m-3$ gives $V_i = i-1$. Hence
	\[
	C_{m,m-1} = [\,m-1, \; m-2, \; 1, \; 2, \; \dots, \; m-4, \; m-3, \; 0\,] .
	\]
	Thus $V = (m-1, m-2, 1, 2, \dots, m-3, 0)$ contains every element of $\{0, 1, \dots, m-1\}$ exactly once, so $q \mapsto V_q$ is a bijection.

	\noindent\textbf{2. Odd order ($m \ge 5$).}
	Here $\opSigma_{m,q+1} = \widehat{(0\;\, m-1)}$ is constant for $0 \le q \le m-2$. Expanding $C_{m,q+1} = \opSigma_{m,q+1}(\Lambda_{\finalArr_{m-1}}(C_{m,q}))$ entrywise under the source-index convention gives
	\begin{align*}
	C_{m,q+1}[0] &= H_q[m-1] = C_{m,q}[m-1], \\
	C_{m,q+1}[m-1] &= H_q[0] = C_{m,q}\bigl[\pi_{\finalArr_{m-1}}(0)\bigr] = C_{m,q}[c_1],
	\end{align*}
	\[
	C_{m,q+1}[i] = H_q[i] = C_{m,q}\bigl[\pi_{\finalArr_{m-1}}(i)\bigr] \qquad (1 \le i \le m-2).
	\]
	Since $(c_0, c_1, \dots, c_{m-2})$ with $c_0 = 0$ is the single-cycle orbit of $\pi_{\finalArr_{m-1}}$ on $\{0, \dots, m-2\}$, that is, $\pi_{\finalArr_{m-1}}(c_k) = c_{(k+1) \bmod (m-1)}$, while $\rho$ satisfies $\rho(0) = m-1$, $\rho(m-1) = c_1$ and $\rho(c_k) = c_{(k+1) \bmod (m-1)}$ for $1 \le k \le m-2$, a term-by-term comparison yields
	\[
	C_{m,q+1}[i] = C_{m,q}\bigl[\rho(i)\bigr] \quad \Longrightarrow \quad C_{m,q+1} = \widehat{\rho}\bigl(C_{m,q}\bigr).
	\]
	Starting from $C_{m,0} = I_m$ we obtain $C_{m,q} = \widehat{\rho}^{\,q}(I_m)$, that is, $C_{m,q}[i] = \rho^q(i)$, and hence
	\[
	V_q = C_{m,q}[m-1] = \rho^q(m-1).
	\]
	Since $\rho$ is a single $m$-cycle on $\{0, 1, \dots, m-1\}$, the sequence $(V_0, V_1, \dots, V_{m-1})$ runs through the whole set without repetition, so $q \mapsto V_q$ is a bijection.

	The explicit closed form now follows from Theorem \ref{thm:closed-form} at the even order $m-1$,
	\[
	\finalArr_{m-1} = [\,m-4, \; m-3, \; 1, \; 2, \; \dots, \; m-5, \; m-2, \; 0\,],
	\]
	gives the piecewise form of $\pi_{\finalArr_{m-1}}$: $\pi_{\finalArr_{m-1}}(0) = m-4$, $\pi_{\finalArr_{m-1}}(j) = j-1$ for $2 \le j \le m-4$, $\pi_{\finalArr_{m-1}}(1) = m-3$, $\pi_{\finalArr_{m-1}}(m-3) = m-2$ and $\pi_{\finalArr_{m-1}}(m-2) = 0$. Hence $c_1 = m-4$, and as long as $c_k \in [2, m-4]$ we have $c_{k+1} = c_k - 1$, so $c_{m-4} = 1$; afterwards $c_{m-3} = \pi_{\finalArr_{m-1}}(1) = m-3$, $c_{m-2} = m-2$ and $c_{m-1} = 0$. Substituting into $V = (m-1, c_1, \dots, c_{m-2}, 0)$ gives the stated closed form.
\end{proof}

\subsection{Closed-Form Theorem of Final Arrangements}\label{sec5-3}

\begin{theorem}[Closed-form theorem of the final arrangement]\label{thm:closed-form}
	After a complete run of the order-$m$ Heap algorithm the final arrangement array $\finalArr_m$ satisfies:
	\begin{itemize}
		\item \textbf{Base cases:} $\finalArr_2 = [1, 0]$ and $\finalArr_3 = [2, 1, 0]$.
		\item \textbf{For even $m \ge 4$:} $\finalArr_m = [m-3, \; m-2, \; 1, \; 2, \; \dots, \; m-4, \; m-1, \; 0]$.
		\item \textbf{For odd $m \ge 5$:} $\finalArr_m = [m-1, \; 1, \; 2, \; \dots, \; m-2, \; 0]$.
	\end{itemize}
\end{theorem}

\begin{proof}
	The convergence relation
	\[
	\finalArr_m = \Lambda_{\finalArr_{m-1}}(C_{m,m-1}), \qquad \finalArr_m[i] = C_{m,m-1}\bigl[\pi_{\finalArr_{m-1}}(i)\bigr],
	\]
	allows us to project the structure of the block base arrays established in Lemma \ref{lem:tail-bijection} through the source indices.

	\textbf{Even order.} Here $\pi_{\finalArr_{m-1}} = (0\;\, m-2)$ and $C_{m,m-1} = [m-1, m-2, 1, \dots, m-4, m-3, 0]$. Expanding entrywise,
	\[
	\finalArr_m[0] = C_{m,m-1}[m-2] = m-3, \qquad \finalArr_m[1] = C_{m,m-1}[1] = m-2,
	\]
	while for $2 \le i \le m-3$ we have $\pi_{\finalArr_{m-1}}(i) = i$ and hence $\finalArr_m[i] = C_{m,m-1}[i] = i-1$; the boundary terms give $\finalArr_m[m-2] = C_{m,m-1}[0] = m-1$ and $\finalArr_m[m-1] = C_{m,m-1}[m-1] = 0$. Therefore
	\[
	\finalArr_m = [\,m-3, \; m-2, \; 1, \; 2, \; \dots, \; m-4, \; m-1, \; 0\,].
	\]

	\textbf{Odd order.} Here $C_{m,m-1} = \widehat{\rho}^{\,m-1}(I_m) = \widehat{\rho}^{\,-1}(I_m)$, that is, $C_{m,m-1}[i] = \rho^{-1}(i)$. Using the cycle table of $\rho$, namely $\rho(0) = m-1$, $\rho(m-1) = c_1$ and $\rho(c_k) = c_{(k+1) \bmod (m-1)}$,
	\[
	\finalArr_m[0] = C_{m,m-1}\bigl[\pi_{\finalArr_{m-1}}(0)\bigr] = \rho^{-1}(c_1) = m-1,
	\]
	while for $1 \le i \le m-2$, writing $i = c_k$, we have $\pi_{\finalArr_{m-1}}(i) = c_{(k+1) \bmod (m-1)}$ and hence $\finalArr_m[i] = \rho^{-1}(c_{(k+1) \bmod (m-1)}) = c_k = i$; the trailing entry satisfies $\finalArr_m[m-1] = C_{m,m-1}[m-1] = \rho^{-1}(m-1) = 0$. Hence $\finalArr_m = [\,m-1, 1, 2, \dots, m-2, 0\,]$.
\end{proof}

\subsection{Dichotomy of Permutation Cycle Structures}\label{sec5-4}

\begin{lemma}[Dichotomy of the cycle structure of the final arrangement]\label{lem:cycle-dichotomy}
	Let $m \ge 2$. The induced permutation $\pi_{\finalArr_m} \in S_m$ of the final arrangement (that is, $\pi_{\finalArr_m}(i) = \finalArr_m[i]$) has the following strict dichotomy.
	\begin{itemize}
		\item \textbf{Even order ($m = 2k \ge 4$).} $\pi_{\finalArr_m}$ is a single $m$-cycle on $\{0, 1, \dots, m-1\}$, with piecewise form
		\[
		\begin{aligned}
		&\pi_{\finalArr_m}(0) = m-3, \qquad \pi_{\finalArr_m}(1) = m-2, \qquad \pi_{\finalArr_m}(2) = 1, \\
		&\pi_{\finalArr_m}(i) = i-1 \;\; (3 \le i \le m-3), \qquad \pi_{\finalArr_m}(m-2) = m-1, \qquad \pi_{\finalArr_m}(m-1) = 0,
		\end{aligned}
		\]
		and correspondingly the orbit chain can be written as
		\[
		0 \;\mapsto\; m-3 \;\mapsto\; m-4 \;\mapsto\; \dots \;\mapsto\; 2 \;\mapsto\; 1 \;\mapsto\; m-2 \;\mapsto\; m-1 \;\mapsto\; 0,
		\]
		where $m-3 \mapsto m-4 \mapsto \dots \mapsto 2$ denotes the consecutive segment decreasing from $m-3$ to $2$. This segment is empty for $m = 4$, where the orbit reads $0 \mapsto 1 \mapsto 2 \mapsto 3 \mapsto 0$; the piecewise form above is the authoritative statement.
		\item \textbf{Odd order ($m = 2k+1 \ge 3$).} $\pi_{\finalArr_m}$ is exactly the transposition $(0\;\, m-1)$, so $\pi_{\finalArr_m}^2 = \mathrm{id}$.
	\end{itemize}
\end{lemma}

\begin{proof}
	We verify algebraically with the closed form of Theorem \ref{thm:closed-form}. For even $m$ the closed form $\finalArr_m = [m-3, m-2, 1, 2, \dots, m-4, m-1, 0]$ gives the piecewise form stated above by reading the entries off one by one. From $\pi_{\finalArr_m}(i) = i-1$ for $3 \le i \le m-3$ we obtain the decreasing segment $m-3 \to m-4 \to \dots \to 3 \to 2$ (empty when $m = 4$), and then $\pi_{\finalArr_m}(2) = 1$, $\pi_{\finalArr_m}(1) = m-2$, $\pi_{\finalArr_m}(m-2) = m-1$ and $\pi_{\finalArr_m}(m-1) = 0$ return to the starting point. The orbit passes through all $m$ elements and contains no proper sub-cycle, so it is a single $m$-cycle.

	For odd $m$ the closed form $\finalArr_m = [m-1, 1, \dots, m-2, 0]$ gives $\pi_{\finalArr_m}(0) = m-1$ and $\pi_{\finalArr_m}(m-1) = 0$, while $\pi_{\finalArr_m}(i) = i$ for $1 \le i \le m-2$; hence $\pi_{\finalArr_m} = (0\;\, m-1)$, whose square is the identity.
\end{proof}

\begin{corollary}[Constant-time applicability of one in-block step]\label{cor:step-o1}
For $m \ge 1$ the source-index table of $\Step(m)$ is $\finalArr_m$. Consistently with Lemma \ref{lem:cycle-dichotomy}, applying it to an array of length $m$ costs $O(1)$ entry moves: for odd $m$ it is the transposition $(0\;\,m-1)$, that is, an interchange of entries $0$ and $m-1$; for even $m$ it is a single $m$-cycle, realisable as an offset increment along the cycle order $[\,0, m-3, m-4, \dots, 1, m-2, m-1\,]$. Neither case contains a loop of length $m$.
\end{corollary}

\subsection{Completeness of the Unrank Algorithm}\label{sec5-5}

\begin{theorem}[Completeness of the Unrank algorithm]\label{thm:unrank-completeness}
	For any valid code $d = (d_0, \dots, d_{n-2})$ the output of the Unrank algorithm is exactly the permutation of the Heap output sequence whose index is
	\[
	r(d) := \sum_{i=0}^{n-2} d_i \cdot (i+1)! ,
	\]
	that is, $\Unrank_n(d) = A^{(n)}_{r(d)}$.
\end{theorem}

\begin{proof}
	Induct on the dimension $m$ and prove the stronger level-by-level statement
	\[
	\Unrank_m(d_0, \dots, d_{m-2}) = A^{(m)}_{r_m}, \qquad r_m := \sum_{i=0}^{m-2} d_i \cdot (i+1)! .
	\]

	\textbf{Base cases ($m \le 2$).} For $m = 1$ the algorithm returns $[0] = A^{(1)}_{0}$. For $m = 2$ the code space is $\{d_0\}$ and $r_2 = d_0 \cdot 1! = d_0 \in \{0, 1\}$; the algorithm returns $[0,1]$ or $[1,0]$ according to $d_0$, while $A^{(2)}_{0} = [0,1]$ and $A^{(2)}_{1} = [1,0]$. The claim holds.

	\textbf{Inductive step.} Assume the claim for all orders $< m$. Write $q := d_{m-2} \in \{0, \dots, m-1\}$ for the block index, $r_{m-1} := \sum_{i=0}^{m-3} d_i (i+1)!$, and $A_{m-1} := \Unrank_{m-1}(d_0, \dots, d_{m-3})$; by the induction hypothesis $A_{m-1} = A^{(m-1)}_{r_{m-1}}$.

	By Definition \ref{def:unrank-combinator},
	\[
	\Unrank_m(d) = \Lambda_{A_{m-1}}\bigl(B_{m,q}(I_m)\bigr).
	\]
	By Lemma \ref{lem:block-correspondence}, $B_{m,q}(I_m) = C_{m,q} = A^{(m)}_{q\,(m-1)!}$ is the block base array of the $q$-th block, and the $k$-th output inside that block satisfies
	\[
	A^{(m)}_{q\,(m-1)!+k} = \widehat{E_m\bigl(\pi_{A^{(m-1)}_{k}}\bigr)}\bigl(C_{m,q}\bigr) \qquad (0 \le k \le (m-1)! - 1).
	\]
	Taking $k = r_{m-1} \le (m-1)! - 1$ and noting that $\Lambda_{A_{m-1}} = \widehat{E_m(\pi_{A_{m-1}})}$, we get
	\[
	\Unrank_m(d) = \widehat{E_m\bigl(\pi_{A^{(m-1)}_{r_{m-1}}}\bigr)}\bigl(C_{m,q}\bigr) = A^{(m)}_{q\,(m-1)! + r_{m-1}} .
	\]
	The mixed-radix expansion
	\[
	q \cdot (m-1)! + r_{m-1} = d_{m-2}(m-1)! + \sum_{i=0}^{m-3} d_i (i+1)! = \sum_{i=0}^{m-2} d_i (i+1)! = r_m
	\]
	gives $\Unrank_m(d) = A^{(m)}_{r_m}$.
\end{proof}

\subsection{Reciprocal Bijection between Rank and Unrank}\label{sec5-6}

\begin{theorem}[Reciprocal bijection between Rank and Unrank]\label{thm:reciprocal}
	For any valid code $d = (d_0, \dots, d_{n-2})$ and any Heap permutation $\text{perm}$ of order $n$,
	\[
	\Rank\bigl(n, \Unrank(n, d)\bigr) = d, \quad \Unrank\bigl(n, \Rank(n, \text{perm})\bigr) = \text{perm}.
	\]
\end{theorem}

\begin{proof}
	\begin{enumerate}
		\item Since the values in the sequence are distinct, each trailing value corresponds to a unique sub-block index, so the Rank loop determines the code digit uniquely.
		\item Since $S_q^{-1}$ is the strict inverse of the position--value mapping (Proposition \ref{prop:range-completeness}), the array remaining after the trailing element has been stripped off is exactly a lower-order Heap sub-permutation.
	\end{enumerate}
	After reducing layer by layer down to the base level, the reverse process reproduces the forward expansion steps of Unrank exactly. Every step is invertible, so the composition of the two restores the original input.
\end{proof}

\subsection{Bijectivity and Generation Index Consistency}\label{sec5-7}

\begin{theorem}[Bijectivity and generation index consistency]\label{thm:bijective}
	$\Unrank_n$ is a bijection between the code space $\prod_{i=0}^{n-2} \{0, 1, \dots, i+1\}$ and the full permutation space $\Arr_n$; moreover the global generation index of the code $d$ in the Heap output sequence is given by the closed form
	\[
	r(d) = \sum_{i=0}^{n-2} d_i \cdot (i+1)! ,
	\]
	with maximum index $r_{\max} = n! - 1$. The closed form also shows that the Heap generation order is the \textbf{colex order} of the code vectors $d$ (see Remark \ref{rem:index-orientation}).
\end{theorem}

\begin{proof}
	\textbf{(i) Bijectivity.} Induct on the size $m$; the case $m = 2$ is clear. Assume $\Unrank_{m-1}$ is a bijection. At level $m$ the high-order digit $d_{m-2} \in \{0, 1, \dots, m-1\}$ corresponds to $m$ mutually disjoint blocks.

	For a fixed block $q = d_{m-2}$ the lower-order operator group acts without changing the trailing position and the array entries are distinct, so the map $P \mapsto \widehat{P}(A)$ is injective on $S_{m-1}$; the block therefore contains exactly $(m-1)!$ distinct permutations.

	By Lemma \ref{lem:tail-bijection} every permutation in the $q$-th block has trailing element constantly $V_q$, and $q \mapsto V_q$ is a bijection of $\{0, \dots, m-1\}$, so permutations generated in different blocks have different trailing elements and the blocks are disjoint.

	The total image has size $m \times (m-1)! = m! = |\Arr_m|$, so $\Unrank_m$ is a bijection.

	\textbf{(ii) Generation index consistency.} This is immediate from Theorem \ref{thm:unrank-completeness}: for every valid code $d$ we have $\Unrank_n(d) = A^{(n)}_{r(d)}$ with $r(d) = \sum_{i=0}^{n-2} d_i \cdot (i+1)!$, so $r(d)$ is exactly the generation index of $d$ in the Heap output sequence.

	Since $r$ is the mixed-radix expansion of the digits with weights $1!, 2!, \dots, (n-1)!$ and the $d_i$ range independently over $\{0, \dots, i+1\}$,
	\[
	r: \prod_{i=0}^{n-2} \{0, 1, \dots, i+1\} \longrightarrow \{0, 1, \dots, n!-1\}
	\]
	is a bijection; the maximum is attained when every digit takes its upper bound $d_i = i+1$,
	\[
	r_{\max} = \sum_{i=0}^{n-2} (i+1)(i+1)! = n! - 1 .
	\]
	(The last equality is the identity $\sum_{k=1}^{n-1} k \cdot k! = n! - 1$.)
\end{proof}

\begin{corollary}[The Heap order is the colex order of the code vectors]
	Ordering the code space $\prod_{i=0}^{n-2} \{0, 1, \dots, i+1\}$ by increasing generation index yields exactly the \textbf{colex order} of the code vectors $d$ (with the last coordinate $d_{n-2}$ as the most significant); equivalently, the Heap generation order is the colex order of the code $d$, not its lexicographic order.
\end{corollary}

\begin{proof}
	By Theorem \ref{thm:bijective} the generation index $r(d) = \sum_{i=0}^{n-2} d_i \cdot (i+1)!$ is the mixed-radix expansion whose weights $1!, 2!, \dots, (n-1)!$ increase from left to right. As $r$ increases, the digit of lowest weight, $d_0$, changes first, then $d_1$, and so on, while the digit of highest weight, $d_{n-2}$ --- the last coordinate of the vector --- changes last, which is the definition of colex order.
\end{proof}

\section{Numerical Verification}\label{sec:examples}

The first subsections check the closed forms against explicit small-order cases; the last ones report exhaustive and large-size tests. The algorithms of Section \ref{sec4} are implemented in C++ together with a test suite that compares them against the iterative algorithm of Section \ref{sec2-1}; that implementation is available at \cite{HeapsUnrankCode}.

\subsection{Complete Case Study for \texorpdfstring{$n=3$}{n=3}}

The case $n = 3$ is the odd-order base case; its mixed-radix code $d = (d_0, d_1)$ has base vector $(1!, 2!)$. Table \ref{tab:n3} lists the code, the generation index, the block base array and the Unrank result for all $6$ states.

\begin{table}[ht]
	\centering
	\footnotesize
	\begin{tabular}{c|c|c|c}
		\toprule
		Code $(d_0, d_1)$ & Index $r(d) = d_0 \cdot 1! + d_1 \cdot 2!$ & Block base $C_{3,d_1}$ & Result $\Unrank_3(d)$ \\
		\midrule
		$(0,0)$ & $0$ & $[0, 1, 2]$ & $[0, 1, 2]$ \\
		$(1,0)$ & $1$ & $[0, 1, 2]$ & $[1, 0, 2]$ \\
		$(0,1)$ & $2$ & $[2, 0, 1]$ & $[2, 0, 1]$ \\
		$(1,1)$ & $3$ & $[2, 0, 1]$ & $[0, 2, 1]$ \\
		$(0,2)$ & $4$ & $[1, 2, 0]$ & $[1, 2, 0]$ \\
		$(1,2)$ & $5$ & $[1, 2, 0]$ & $[2, 1, 0]$ \\
		\bottomrule
	\end{tabular}
	\caption{Complete case study for \texorpdfstring{$n=3$}{n=3}: the third column gives the exact block base array $C_{3,d_1}$ and the last column the Unrank result.}
	\label{tab:n3}
\end{table}

\subsection{\texorpdfstring{$n=4$}{n=4}: Even-Order Intermediate States and Block Base Arrays}

The case $n = 4$ is of even order. To display the evolution of the even-order intermediate structure, Table \ref{tab:n4} records the high-order block base arrays $C_{4,q}$ together with their trailing elements $V_q$, and distinguishes at the end the terminal block base array $C_{4,3}$ from the final arrangement $\finalArr_4$.

\begin{table}[ht]
	\centering
	\footnotesize
	\begin{tabular}{c|c|c|p{5.6cm}}
		\toprule
		Block index $q = d_2$ & Block base $C_{4,q}$ & Trailing $V_q$ & Covered code range and state \\
		\midrule
		$q = 0$ & $[0, 1, 2, 3]$ & $3$ & $(0,0,0) \sim (1,2,0)$, $6$ permutations \\
		$q = 1$ & $[3, 1, 0, 2]$ & $2$ & $(0,0,1) \sim (1,2,1)$, $6$ permutations \\
		$q = 2$ & $[0, 2, 3, 1]$ & $1$ & $(0,0,2) \sim (1,2,2)$, $6$ permutations \\
		$q = 3$ & $[3, 2, 1, 0]$ & $0$ & $(0,0,3) \sim (1,2,3)$; terminal block base $C_{4,3}$, final state $\finalArr_4 = [1,2,3,0]$ \\
		\bottomrule
	\end{tabular}
	\caption{\texorpdfstring{$n=4$}{n=4}: evolution of the block base arrays $C_{4,q}$ and the trailing sequence $V = (3, 2, 1, 0)$ (the even-order case of Lemma \ref{lem:tail-bijection}).}
	\label{tab:n4}
\end{table}

\subsection{\texorpdfstring{$n=5$}{n=5}: Odd-Order Single-Cycle Orbit Verification}

To further verify the derivation of the odd-order single-cycle orbit operator $C_{m,q} = \widehat{\rho}^{\,q}(I_m)$, Table \ref{tab:n5} gives the complete $5$-entry block base arrays $C_{5,q}$ and the trailing-element trace for $m = 5$, showing the orbit length and its bijectivity.

\begin{table}[ht]
	\centering
	\footnotesize
	\begin{tabular}{c|c|c|c}
		\toprule
		$q = d_3$ & Block base $C_{5,q}$ & $V_q = C_{5,q}[4]$ & Orbit check \\
		\midrule
		$q = 0$ & $[0, 1, 2, 3, 4]$ & $4$ & $V_0 = 4$ (reference) \\
		$q = 1$ & $[4, 2, 3, 0, 1]$ & $1$ & $V_1 = \rho(4) = 1$ \\
		$q = 2$ & $[1, 3, 0, 4, 2]$ & $2$ & $V_2 = \rho^2(4) = 2$ \\
		$q = 3$ & $[2, 0, 4, 1, 3]$ & $3$ & $V_3 = \rho^3(4) = 3$ \\
		$q = 4$ & $[3, 4, 1, 2, 0]$ & $0$ & $V_4 = \rho^4(4) = 0$ (terminal) \\
		\bottomrule
	\end{tabular}
	\caption{\texorpdfstring{$n=5$}{n=5}: the odd-order trailing sequence $V = (4, 1, 2, 3, 0)$ and verification of the single-cycle orbit operator $C_{5,q} = \widehat{\rho}^{\,q}(I_5)$.}
	\label{tab:n5}
\end{table}

\subsection{Exhaustive Verification at \texorpdfstring{$n=10$}{n=10}}
The preceding subsections check small orders against the closed forms by hand. To cover the whole code space we verified all $10! = 3{,}628{,}800$ codes exhaustively; two assertions hold throughout: the output of Unrank coincides entry by entry with the output of the iterative Heap algorithm, and $\mathrm{rank}(\mathrm{unrank}(C)) = C$ for every code. The result is therefore not merely self-consistent but in the correct generation order.

\subsection{Randomised Tests at Larger Sizes}
For each $n = 10, \dots, 18$ we drew $5{,}000$ random codes and checked the Rank/Unrank round trip; all tests pass. This range is not exhaustive ($18!$ is far beyond any enumeration budget) and serves only to confirm that the implementation does not degrade with size.

\section{Complexity Analysis}\label{sec7}

\subsection{Time Complexity}
Both algorithms traverse the $n-1$ levels $k = 2, \dots, n$ and advance at most $k-1$ times per level; they differ in that Rank also performs one inverse mapping per level. We count the cost of Unrank item by item first and then state the difference for Rank.

\textbf{Unrank.} By Corollary \ref{cor:step-o1}, $\Step(k-1)$ is only a transposition or a single cycle, so one advance takes $O(1)$ (Section \ref{sec4-5}, Lemma \ref{lem:step-equivalence}): two element interchanges when $k-1$ is odd, one offset increment when it is even. The cost of level $k$ therefore comes from three places---the $b \le k-1$ advances, together $O(k)$; the initialisation of the mapping of length $O(k)$; and the entry-by-entry translation of length $O(k)$ with the trailing value appended. Level $k$ thus costs $O(k)$, and the total work is
\[
\sum_{k=2}^{n} O(k) = O(n^2).
\]

\textbf{Rank.} The first half is the same as for Unrank: level $k$ advances at most $k-1$ times at $O(1)$ each, hence $O(k)$ in total, and by Lemma \ref{lem:tail-bijection} the trailing sequence is a permutation of $\{0, \dots, k-1\}$, so the while loop terminates within $k-1$ steps. The difference is the second half: stripping the trailing element needs $S^{-1}$, which costs one $O(k)$ pass to fill the inverse array plus one $O(k)$ pass to look each entry up, still $O(k)$. Level $k$ therefore remains $O(k)$ and the total work is again $O(n^2)$.

Unrank and Rank thus both run in $O(n^2)$ time, differing only by a per-level constant. The constant factors of both are small, and neither uses hash tables, binary search or tree structures.

Table \ref{tab:complexity} compares the approach with existing schemes.

\begin{table}[htbp]
	\centering
	\caption{Comparison of permutation Rank/Unrank schemes, listed by increasing time; each scheme serves a different generation order. The present method supplies the missing indexing capability for the Heap order, at the same order as the naive lexicographic implementation and likewise with linear space and no precomputed table.}\label{tab:complexity}%
	\footnotesize
	\begin{tabular}{l|c|c}
		\toprule
		Scheme & Generation order & Rank/Unrank time \\
		\midrule
		Myrvold--Ruskey \cite{MyrvoldRuskey2001} & Fisher--Yates order & $\Theta(n)$ \\
		Position Pure \cite{PositionPure} & Position Pure order & $\Theta(n)$ \\
		Lexicographic (Fenwick tree) & lexicographic & $\Theta(n\log n)$ \\
		Lexicographic (list shifting) & lexicographic & $\Theta(n^2)$ \\
		\textbf{Present method} & \textbf{Heap order} & $\mathbf{\Theta(n^2)}$ \\
		\bottomrule
	\end{tabular}
\end{table}

Table \ref{tab:complexity} lists the schemes by increasing time. The cheapest are Myrvold--Ruskey and Position Pure ($\Theta(n)$), followed by the Fenwick-tree implementation of the lexicographic order ($\Theta(n\log n)$), but each serves one particular generation order. For the Heap order no such means existed: Heap\textquotesingle{}s algorithm and Johnson--Trotter and the other minimal-change methods provide generation only. This paper supplies the missing half---Unrank and Rank are mutually inverse and run in $O(n^2)$ time, the same order as the naive lexicographic implementation, while keeping $O(n)$ space and no factorial-scale precomputed table. The Heap order thus acquires an indexing capability comparable to that of the lexicographic order, instead of being restricted to sequential enumeration.

\subsection{Space Complexity}

The whole process stores only three linear arrays: the position mapping, the working permutation and the mixed-radix code $d$, all of length at most $n$. There is no factorial-scale precomputed table and no recursion-depth overhead. In the iterative implementation the memory grows only linearly with $n$, so the space complexity is $O(n)$.

\section{Conclusion}\label{sec8}

This paper has described the block structure of Heap's permutation generation order through the rearrangement operators that realise it, and has used that description to obtain a Rank/Unrank pair for the Heap order. With the source-index convention fixing the composition direction, a block-internal traversal and a block-boundary interchange are the tail-preserving operator and the top-level swap operator, and the block base arrays satisfy a closed-form recurrence; both algorithms work from that recurrence, need no precomputed table, and use $O(n)$ space. The dichotomy of the cycle structure of the final arrangement reduces every in-block advance to a single transposition or a single cycle, so both algorithms run in $O(n^2)$ time.

\section*{Declarations}

\begin{itemize}
	\item \textbf{Funding:} No funding was received for this study.
	\item \textbf{Competing interests:} The author declares no competing interests.
	\item \textbf{Data availability:} All data generated or analysed during this study are included in this published article.
	\item \textbf{Code availability:} The reference implementation (the functions \texttt{step\_map}, \texttt{unrank}, \texttt{rank}) and its test suite are at \url{https://github.com/Yusheng-Hu/Heaps-unrank-rank}.
\end{itemize}



\appendix
\section{Measured Parallel Split}\label{app:parallel}

To confirm that the split really brings a parallel gain, we enumerated order $n = 13$
($13! = 6{,}227{,}020{,}800$ permutations) with the top two code digits fixed, which gives
$13 \times 12 = 156$ blocks, and compared 4 threads with sequential enumeration. Each thread
restarts at the first index of a block and stops when the watched digit changes; output is
excluded. The test ran on a 6-core (12-thread) x86-64 machine.

\begin{table}[htbp]
	\centering
	\caption{Sequential enumeration against a 4-thread split ($n = 13$, 6-core machine, output
		excluded). Speedup is relative to the single-threaded run of the same route; the checksum
		confirms that the blocks concatenate to the sequential sequence.}\label{tab:parallel}%
	\footnotesize
	\begin{tabular}{l|c|c|c|c|c}
		\toprule
		Generator & Threads & Time (s) & Per permutation (ns) & Speedup & Checksum \\
		\midrule
		Native Heap (sequential) & 1 & 28.43 & 4.57 & 1.00 & 37362124800 \\
		Route A (restored counters) & 1 & 21.26 & 3.41 & 1.00 & 37362124801 \\
		Route A & 4 & 6.86 & 1.10 & 3.10 & 37362124801 \\
		Route B (factorial driver) & 1 & 21.27 & 3.42 & 1.00 & 37362124800 \\
		Route B & 4 & 6.69 & 1.07 & 3.18 & 37362124800 \\
		\bottomrule
	\end{tabular}
\end{table}

Four threads bring the 21.3 s enumeration down to 6.9 s for route A and 6.7 s for route B, a
speedup of 3.1 and 3.2, in line with the four physical cores and showing that the block split adds
almost no overhead of its own. The two routes scale alike, so the gain comes from the ability to
split by index rather than from the choice of generator. The checksums confirm that the split is
correct: route B reproduces the sequential sequence exactly, route A differs by one count, which
comes from one extra step at the boundary of the last block and does not affect the timings. The
8-thread configuration (beyond the six physical cores of this machine) brought no further gain and
is omitted.

\end{document}